\documentclass[prl,aps,superscriptaddress,twocolumn,10pt,nolongbibliography]{revtex4-2}
\usepackage[pdftex]{graphicx}
\usepackage{color}
\usepackage{amsmath,amsthm,amssymb,mathtools}
\usepackage[colorlinks=true,allcolors=blue]{hyperref}
\usepackage[export]{adjustbox}
\usepackage{soul}

\usepackage{listings}
\usepackage{color}
\definecolor{codegreen}{rgb}{0,0.6,0}
\definecolor{codegray}{rgb}{0.5,0.5,0.5}
\definecolor{codepurple}{rgb}{0.58,0,0.82}
\definecolor{backcolour}{rgb}{0.95,0.95,0.92}
\lstdefinestyle{mystyle}{
    backgroundcolor=\color{backcolour},
    commentstyle=\color{codegreen},
    keywordstyle=\color{magenta},
    numberstyle=\tiny\color{codegray},
    stringstyle=\color{codepurple},
    basicstyle=\footnotesize,
    breakatwhitespace=false,
    breaklines=true,
    captionpos=b,
    keepspaces=true,
    numbers=left,
    numbersep=5pt,
    showspaces=false,
    showstringspaces=false,
    showtabs=false,
    tabsize=2
}
\begin{document}
\title{Registry-dependent potential energy and lattice corrugation of twisted bilayer graphene from quantum Monte Carlo}
\author{Kittithat Krongchon}
\affiliation{Department of Physics, Institute for Condensed Matter Theory, University of Illinois at Urbana-Champaign}
\author{Tawfiqur Rakib}
\affiliation{Department of Mechanical Science and Engineering, University of Illinois at Urbana-Champaign}
\author{Shivesh Pathak}
\affiliation{Sandia National Laboratories}
\author{Elif Ertekin}
\affiliation{Department of Mechanical Science and Engineering, University of Illinois at Urbana-Champaign}
\affiliation{Materials Research Laboratory, University of Illinois at Urbana-Champaign}
\author{Harley T. Johnson}
\affiliation{Department of Mechanical Science and Engineering, University of Illinois at Urbana-Champaign}
\affiliation{Department of Materials Science and Engineering, University of Illinois at Urbana-Champaign}
\author{Lucas K. Wagner}
\affiliation{Department of Physics, Institute for Condensed Matter Theory, University of Illinois at Urbana-Champaign}
\date{\today}

\begin{abstract}
An uncertainty in studying twisted bilayer graphene (TBG) is the minimum energy geometry, which strongly affects the electronic structure.
The minimum energy geometry is determined by the potential energy surface, which is dominated by van der Waals (vdW) interactions.
In this work, large-scale diffusion quantum Monte Carlo (QMC) simulations are performed to evaluate the energy of bilayer graphene at various interlayer distances for four stacking registries. 
An accurate registry-dependent potential is fit to the QMC data and is used to describe interlayer interactions in the geometry of near-magic-angle TBG.
The band structure for the optimized geometry is evaluated using the accurate local-environment tight-binding model.
We find that compared to QMC, DFT-based vdW interactions can result in errors in the corrugation magnitude by a factor of 2 or more near the magic angle. 
The error in corrugation then propagates to the flat bands in twisted bilayer graphene, where the error in corrugation can affect the bandwidth by about 30\% and can change the nature and degeneracy of the flat bands.
\end{abstract}

\maketitle
\section{Introduction}

Twisted bilayer graphene (TBG) exhibits a multitude of correlated electronic phases and has emerged as a platform for studying correlated electron physics~\cite{bistritzer2011moire,cao2018unconventional,cao2018correlated,naik2018ultraflatbands,padhi2018doped,choi2019electronic,shen2020correlated,basov2021polariton,padhi2021generalized}.
These correlation-driven phases are attributed to flattening of the band structure near the Fermi level due to the superlattice interaction in moir\'e patterns~\cite{dean2013hofstadter,liu2014evolution,yeh2016direct,carr2017twistronics,ribeiro2018twistable,lisi2021observation}.
However, a complete model Hamiltonian for these systems remains unknown.
An important piece of the puzzle of predicting flat bands is the van der Waals (vdW) interactions, which lead to symmetry breaking, corrugation, and other distortions in the layers.
The lattice reconstruction in the moir\'e superlattices in turn significantly affects the electronic behavior of this system~\cite{uchida2014atomic,dai2016twisted,nam2017lattice,yoo2019atomic,pathak2022accurate,rakib2022corrugation,zhou2023coexistence}.
Therefore, the interplay between lattice corrugation and electronic structure provides strong impetus to accurately model vdW contributions in TBG.

Evaluating vdW interactions, however, is a challenging task for two reasons. 
First, this type of interaction results from long-range electron correlations, which means that local or semi-local exchange-correlation functionals from density functional theory (DFT)  cannot describe them~\cite{allen2002helium,zimmerli2004dispersion,tsuzuki2001interaction}. 
Although a large set of computational methods has been developed in the DFT regime to account for the long-range interactions~\cite{grimme2006semiempirical,barone2009role,grimme2010consistent}, they are empirical in nature. 
Within various iterations of the so-called DFT-D scheme, which adds the dispersion corrections to the standard Kohn--Sham DFT energy, the estimated binding energy errors can differ by up to 250\%~\cite{mostaani2015quantum} in bilayer graphene. 
Thus, the large uncertainty in this set of techniques warrants a more accurate first-principles approach. 
The second reason for the difficulty is that accurate treatment of vdW interactions becomes computationally intractable for a large system, such as small-angle TBG, which consists of $\sim 10^4$ atoms. 
A systematic potential training approach is needed to model lattice and electronic degrees of freedom for vdW systems, while maintaining first-principles accuracy.

Diffusion quantum Monte Carlo (QMC) has been shown to closely reproduce experimental values for vdW materials due to explicit treatment of electron interactions~\cite{ganesh2014binding,benali2014application,shin2017nature,krogel2020perspectives}.
Specifically, in AB-stacked bilayer graphene, the binding-energy curve from QMC is able to predict the out-of-plane phonon frequency and relaxed interlayer spacing in agreement with available experimental results~\cite{mostaani2015quantum}, which shows that QMC is a promising technique for investigating bilayer graphene.
The interlayer energy curve from QMC data for this system is available for only a single registry: AB-stacked bilayer graphene. 
To fit a registry-dependent potential, multiple registries are needed.
Therefore, more QMC data is needed to accurately parameterize the potential energy surface of the entire moir\'e superlattice.

In this manuscript, we use large-scale QMC simulations to compute a reference quality Born--Oppenheimer ground state energy for bilayer graphene as a function of the displacement for four stacking registries, which allows for an accurate assessment of the stacking fault energy.
To make this high quality potential usable for large-scale molecular dynamics calculation, we fit the QMC results to an interatomic potential.
We find that the QMC data is closer to the random-phase approximation (RPA)~\cite{zhou2015van} and a previous atomic potential parameterization~\cite{ouyang2018nanoserpents} than  to DFT with dispersion corrections. 
The data is well-fit by the Kolmogorov--Crespi (KC) potential model~\cite{kolmogorov2005registry}, whose refined parameters for the KC potential are made available in the supplementary information in \lstinline{LAMMPS}~\cite{thompson2022lammps,plimpton1995fast} format.
We then assess the importance of an accurate vdW energy calculation by investigating the effect of corrugation computed by QMC, DFT-D2~\cite{grimme2006semiempirical,barone2009role}, DFT-D3~\cite{grimme2010consistent}, and Ref.~\cite{ouyang2018nanoserpents} (labeled as ``KC-Ouyang'').
For each relaxed structure from different parameterizations of the KC potential, the band structure is evaluated using the accurate local-environment tight-binding (LETB) model~\cite{pathak2022accurate}.
The DFT-D2 and DFT-D3 vdW interactions result in large errors in the band structure due to the poor estimation of the structural corrugation.
We show that the difference in corrugation due to QMC evaluation of the vdW interaction leads to significant changes in the electronic band structure.

\section{Method}

\begin{figure*}
\includegraphics[width=7in]{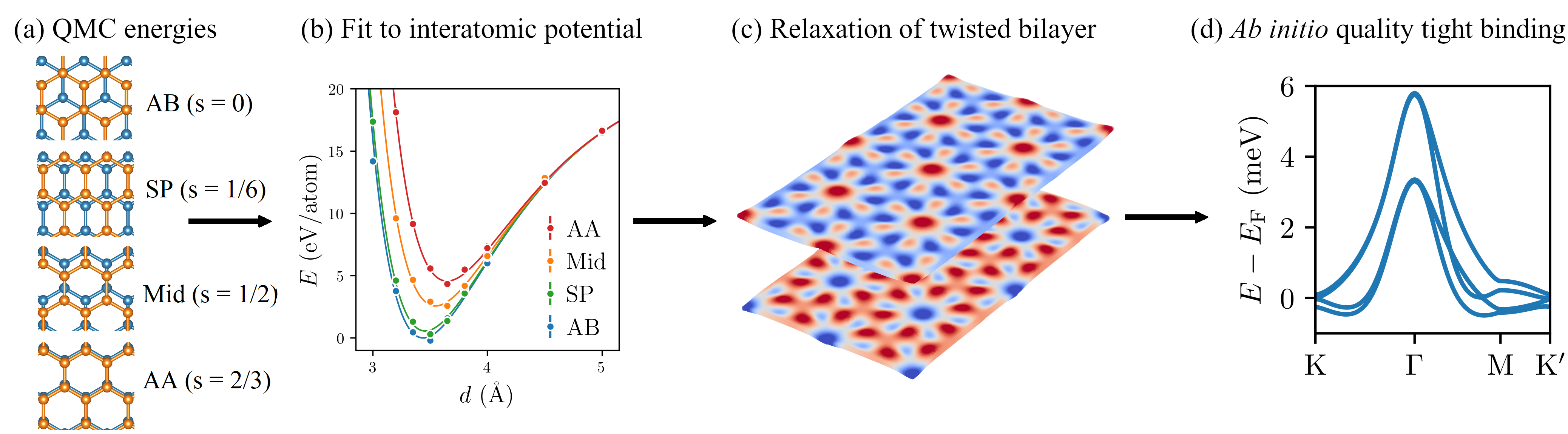}
\caption{
Workflow for training the KC potential and calculating the band structure. 
(a) First, the QMC energy for bilayer graphene is calculated at various interlayer distances for four stacking registries. 
Stacking registries are labeled according to the rigid translation distance $s$ of the top graphene layer along the armchair direction in units of $\sqrt{3} a$, where $a = 2.46~\text{\AA}$ is the in-plane lattice constant.
(b) In the second step, the KC potential is fit to the QMC data. The error bars represent the single $\sigma$ statistical uncertainties associated with the QMC results, which are explained in detail in the Appendix.
(c) Next, using the fitted KC interlayer interaction, the TBG structure is relaxed. 
(d) Lastly, the band structure is computed on the minimum energy structure.
}
\label{fig:workflow}
\end{figure*}

The procedure for parameterizing the interlayer interaction of TBG and evaluating the sensitivity of the minimum energy structure and electronic structure is outlined as follows.
\begin{enumerate}
    \item First, the energy of rigid bilayer graphene as a function of the interlayer distance, is sampled using QMC for four stacking registries, defined in Fig.~\ref{fig:workflow}.
    \item The sampled QMC energy data are used to fit the KC potential.
    \item Using the fitted KC potential, the structure of TBG is optimized by minimizing the total energy.
    \item The electronic structure is calculated on the relaxed structure of TBG using the LETB model~\cite{pathak2022accurate}.
\end{enumerate}
The steps are illustrated in Fig.~\ref{fig:workflow} and are discussed in further details in the following sections.

\subsection{QMC calculations}
We consider 4 stacking registries of bilayer graphene, which are constructed by performing rigid translations of the top layer of AB-stacked bilayer graphene along the armchair direction by different distances.
We label these registries according to the translation distances $s$ as listed in~\ref{fig:workflow}.
The sliding parameter $s$ is defined such that starting from the AB structure ($s = 0$), the translation of $s = 1$ brings the structure back to AB. 
The nomenclature of these stacking orders is described in Ref.~\cite{zhou2015van}, except for the ``Mid'' stacking type, which is additionally defined here to be $s = 1/2$.
For each of the four registries, we perform QMC to sample the energy at 11 interlayer distances, ranging from $d = 3~\text{\AA}$ to $7~\text{\AA}$, for a total of 44 energy data points.

In the fixed-node QMC scheme as implemented in the \lstinline{QMCPACK} software package~\cite{kim2018qmcpack}, the ground-state wave function is projected out of the Slater--Jastrow trial wave function of the form
\begin{align}
\Psi(\mathbf{R}) &= \mathrm{Det} [\phi_i^{\uparrow}(\mathbf{r}_j^{\uparrow})] \mathrm{Det} [\phi_i^{\downarrow}(\mathbf{r}_j^{\downarrow})] \exp(J),
\end{align}
where $\mathbf{R} = \{\mathbf{r}_1, \ldots, \mathbf{r}_M\}$ is the collection of electron coordinates of the $M$-electron system, $\phi$ is the Kohn--Sham orbital, $i$ and $j$ denote electron indices, $\uparrow$ and $\downarrow$ indicate spins, and $J$ is the Jastrow correlation factor as defined in Ref.~\cite{mitavs1994quantum}.

The set of Kohn--Sham orbitals is produced by the \lstinline{Quantum ESPRESSO} plane-wave DFT code~\cite{giannozzi2009quantum,giannozzi2017advanced,giannozzi2020quantum}, using 200 Ry kinetic energy cutoff and the $\mathbf{k}$-point mesh of $20 \times 20 \times 1$.
The core electrons are removed using Dirac--Fock pseudopotentials described by Ref.~\cite{trail2005norm,trail2005smooth}.
We verify that the orbitals generated by the Perdew--Burke--Ernzerhof approximation~\cite{perdew1996generalized} and two vdW functionals, namely DFT-D2~\cite{grimme2006semiempirical} and DFT-D3~\cite{grimme2010consistent}, result in the same QMC energy within error bars~\cite{krongchon2017accurate}.
The QMC energy is twist-averaged~\cite{lin2001twist} over $4 \times 4$ $\mathbf{k}$-point mesh.
The finite-size errors are eliminated by extrapolating the twist-averaged QMC energies of $3 \times 3$, $4 \times 4$, $5 \times 5$, and $6 \times 6$ bilayer graphene supercells to the thermodynamic limit.
The full analysis of the QMC errors is provided in the Appendix.

\subsection{Potential parameterization}
The stacking-dependent KC potential~\cite{kolmogorov2005registry} is fit using least squares to the 44 calculated QMC data points to obtain a smooth curve that can describe the energy at any registry and interlayer distance. 
Since the KC potential does not provide a good fit across the entire range of interlayer distances, we weight the low-energy configurations higher than the large-distance configurations. 
In this work, we assume that this is enough to estimate the interlayer interaction in the twisted bilayer case; essentially we are assuming that the interaction energy only depends on the local stacking. 
We thus expect the potential to be most accurate for small twist angles, where the local alignment of the layers varies slowly.
The detailed procedure to select the most suitable model for a low energy structure is described in the Appendix.  

The KC potential model fit to QMC data is labeled as KC-QMC.
We perform a similar procedure for DFT-D2~\cite{grimme2006semiempirical} and DFT-D3~\cite{grimme2010consistent} but with many more samples of interlayer distances ($0.01~\text{\AA}$ apart) as these vdW correction schemes are not computationally expensive. 
The KC potentials fitted to DFT-D2 and DFT-D3 are labeled as KC-DFT-D2 and KC-DFT-D3 respectively in order to make distinctions between the computed training data and the fitted curves.

\subsection{Stacking-fault energy calculation}
The stacking-fault energy (SFE) is defined as the difference between the energy of a displaced configuration ($s \neq 0$) and the energy of the most stable configuration ($s = 0$), which is the AB-stacked bilayer graphene in this case.
Given a KC potential, we find the minimum point of the energy curve for each registry to obtain the relaxed interlayer spacing and its corresponding minimum energy.
We plot the SFE as a function of registry $s$ along the armchair direction for our results from KC-QMC, KC-DFT-D2, and KC-DFT-D3 in Fig.~\ref{fig:gsfe}(a).
The KC parameters for KC-QMC (Table~\ref{tab:kc_params}), KC-DFT-D2, and KC-DFT-D3 are obtained by the fitting procedure as described in the Appendix, while the KC parameters for KC-Ouyang are taken from Ref.~\cite{ouyang2018nanoserpents}.
The data points for KC-QMC, KC-Ouyang, KC-DFT-D2, and KC-DFT-D3 are obtained from a quadratic fit within the range of 0.05~\text{\AA} from the minimum of the potential energy surface for each stacking registry.
The data points for RPA are taken directly from Ref.~\cite{zhou2015van}.

We fit the SFE as a function of registry using the formula~\cite{zhou2015van}
\begin{align}
F(s) = c_0
&+ c_1 \left[ 2\cos(2\pi s) + \cos(4\pi s) \right] \nonumber \\
&+ c_2 \left[ 2\cos(6\pi s) + 1 \right] \nonumber \\
&+ c_3 \left[ 2\cos(4\pi s) + \cos(8\pi s) \right] \nonumber \\
&+ \sqrt{3} c_1 \left[ -2\sin(2\pi s) + \sin(4\pi s) \right] \nonumber \\
&- \sqrt{3} c_3 \left[ -2\sin(4\pi s) + \sin(8\pi s) \right], \label{eq:gsfe}
\end{align}
where $F(s)$ is the SFE, $s$ is the registry in the armchair direction in units of $\sqrt{3} a$, and $a$ is the in-plane lattice constant.
The fitting constants $c_0$, $c_1$, $c_2$, $c_3$ are reported in Table~\ref{tab:gsfe_constants}.
We also fit the interlayer distance $d_{\textrm{min}}$ for a given registry using the same functional form as Eq.~(\ref{eq:gsfe}).

\subsection{Structural relaxation}
The relaxed geometries of TBG are obtained using the conjugate gradient method with a stopping tolerance of $10^{-11}~\mathrm{eV}$ as implemented in the \lstinline{LAMMPS} molecular dynamics program.
The intralayer interaction is given by the reactive empirical bond order potential~\cite{brenner2002second}, while the interlayer interaction is described by one of the four KC potentials that we have parameterized, namely KC-QMC, KC-Ouyang, KC-DFT-D2, and KC-DFT-D3.
The initial structures are defined by two rigid sheets of graphene at $3.4~\text{\AA}$ interlayer distance. 
The top layer is rotated at a twist angle $\theta$ with respect to the bottom layer.
For this work, we perform relaxation calculations for the commensurate twist angles of $\theta = 0.84^{\circ}, 0.93^{\circ}, 0.99^{\circ}, 1.05^{\circ}, 1.08^{\circ}$, and $1.16^{\circ}$.
Despite the large number of atoms in a simulation cell, which is on the order of $10^4$, the geometry optimizations remain computationally tractable due to the low cost of the classical potentials.

\subsection{Local-environment tight-binding model}
To determine the band structure of the relaxed geometry, the local-environment tight-binding model is employed due to its high accuracy within twisted bilayer graphene~\cite{pathak2022accurate}. 
This model accounts for the detailed local environment of the nearby atoms and has the following form.
\begin{align}
H_{\textrm{LETB}} &= \sum_{ij \sigma} t_{ij}^{\textrm{LETB}} (\mathbf{R}_i, \mathbf{R}_j, \{ \mathbf{R}_{ij}\}) c_{i \sigma}^{\dagger} c_{j \sigma} + \textrm{h.c.},
\end{align}
where $\sigma$ denotes the spin index, $\mathbf{R}_i$ and $\mathbf{R}_j$ represent the location of atoms $i$ and $j$, $\{\mathbf{R}_{ij}\}$ is a set of atomic positions in the vicinity of atoms $i$ and $j$. 
The functional form of the hopping parameters $t_{ij}^{\textrm{LETB}}$ is classified into intralayer and interlayer contributions based on the $z$ projection of the distance $\mathbf{R}_i - \mathbf{R}_j$. The detailed description of the functional form of $t_{ij}^{\textrm{LETB}}$ is provided in Ref.~\cite{pathak2022accurate}.

\section{Results}

\begin{figure}
    \includegraphics{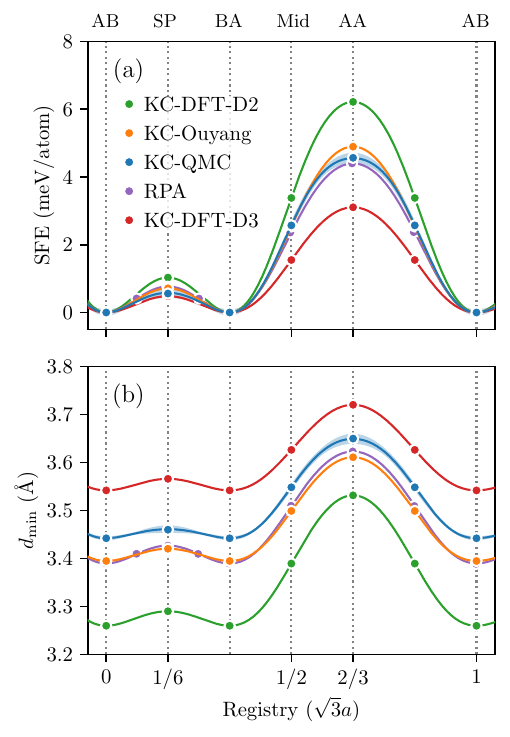}
    \caption{
    (a) Stacking-fault energy given by different potential energy surfaces. Data points of KC-QMC, KC-Ouyang, KC-DFT-D2, and KC-DFT-D3 are obtained from quadratic fits in the range $0.05~\text{\AA}$ around the potential energy surface minimum. The RPA data points are obtained from Ref.~\cite{zhou2015van}. Lines are fits of Eq.~(\ref{eq:gsfe}) to the data points. QMC errors from the bootstrapping technique are represented by the light blue regions around the QMC fit.
    (b) Minimum interlayer spacings. Data points are obtained from the same procedure as the stacking-fault energy.
    }
    \label{fig:gsfe}
\end{figure}

The results of our calculations are discussed using QMC as the reference.
This decision is based on the ability to reproduce the experimental values of relaxed interlayer spacing in the AB stacking and out-of-plane zone-center optical phonon frequency in bilayer graphene~\cite{mostaani2015quantum}.

\subsection{Stacking-fault energy}

Figure~\ref{fig:gsfe}(a) shows the SFE as a function of registry.
We find a close agreement between KC-QMC, KC-Ouyang, and RPA.
Let us consider the AA stacking, shown by the vertical dotted line at the registry of $s = 2/3$, where the differences are the most notable.
The difference between KC-QMC and RPA is 0.2~meV in the stacking-fault energy.
We consider the agreement between KC-QMC and RPA to be an indication that both might be accurate in this case.
A similar agreement between QMC and RPA was found for water on boron nitride~\cite{wu2016hexagonal}.
The difference between KC-QMC and KC-Ouyang is approximately 0.3~meV in the stacking-fault energy.
While the agreement might appear to be fortuitous, a similar agreement between QMC and the so-called DFT-MBD~\cite{ambrosetti2014long}, on which the KC-Ouyang potential was trained, was found in DNA-ellipticine molecules~\cite{benali2014application}.

Relative to the consensus SFE of KC-QMC, RPA, and KC-Ouyang, KC-DFT-D2 overestimates SFE by 1.7~meV for the AA stacking compared to KC-QMC, while KC-DFT-D3 underestimates the SFE by 1.5~meV. 
This result agrees with previous comparisons to RPA only~\cite{zhou2015van}, so our results increase confidence in the RPA results.

Figure~\ref{fig:gsfe}(b) shows the relaxed interlayer distance, $d_{\textrm{min}}$, as a function of registry.
Similar to the case of SFE in Fig.~\ref{fig:gsfe}(a), there is an agreement between KC-QMC, KC-Ouyang, and RPA. 
In this case, however, both RPA and KC-Ouyang underestimate the relaxed interlayer distance for the AA stacking by a small amount of $0.03$ and $0.04~\text{\AA}$, respectively.
Meanwhile, the two dispersion-corrected DFT methods show the opposite trend of SFE, as shown by the fact that KC-DFT-D2 has the smallest relaxed interlayer distances overall. 
KC-DFT-D2 underestimates the relaxed interlayer distance by $0.12~\text{\AA}$, while KC-DFT-D3 overestimates it by $0.07~\text{\AA}$.
This result is consistent with the result of SFE shown in Fig.~\ref{fig:gsfe}(a) in the sense that a smaller interaction leads to larger interlayer distance which in turn results in smaller SFE.

\subsection{Effects of the more accurate interlayer potential on the minimum energy structure of TBG}
\begin{figure*}
~~~~~~~~~~~\includegraphics[width=7in]{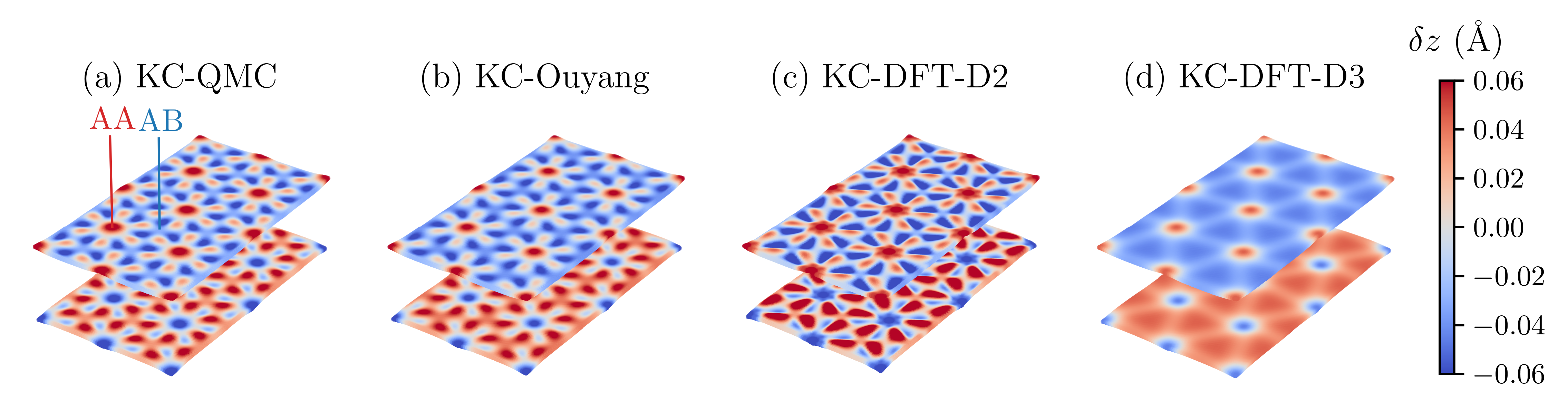}
\includegraphics[width=6.7in,left]{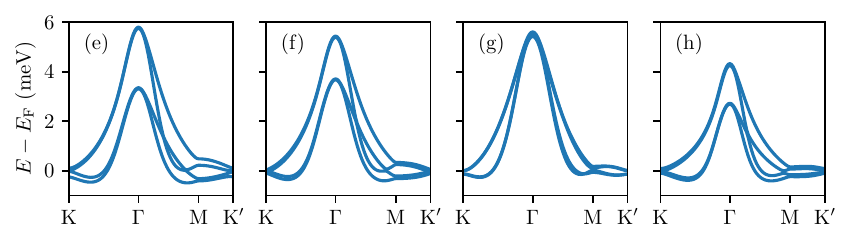}
\caption{
Corrugation of twisted bilayer graphene at $0.99^{\circ}$ from various sources and their respective flat bands.
(a, b, c, d) Corrugation in angle twisted bilayer graphene geometry, relaxed using KC potentials with parameter sets from difference sources---(a) QMC, (b) Ref.~\cite{ouyang2018nanoserpents}, (c) DFT-D2, and (d) DFT-D3. 
Corrugation, $\delta z$, is defined as the deviation of the $z$-coordinate of a carbon atom from the average of the smallest and largest $z$-coordinates within the layer. 
(e, f, g, h) Flat bands near the Fermi level for four twisted bilayer structures obtained from different potentials as described in (a, b, c, d).
}
\label{fig:0-99}
\end{figure*}

The corrugation $\delta z$, defined as the deviation of the $z$-coordinate of a carbon atom from the average of the smallest and largest $z$-coordinates within the layer, of both layers for each interlayer potential is visualized in Fig.~\ref{fig:0-99}(a, b, c, d).
In the AA regions, the relaxed structures manifest an upward bulge in the top layer and a downward bulge in the bottom layer, which are denoted by the most prominent red and blue regions in the visualization.
Around these AA peaks, the 6-fold symmetric structure is observed. 
The formation of small-amplitude structure around the AA regions results in the 3-fold symmetry around the AB/BA regions, which can be identified by the centroid of three adjacent AA nodes as labeled in Fig.~\ref{fig:0-99}(a).
This result is in qualitative agreement with previous studies of breathing mode structure~\cite{dai2016twisted,dai2016structure,rakib2022corrugation}.
The heights of these peaks depend on the potential being used to describe the interlayer interactions. 
In the case of KC-QMC (Fig.~\ref{fig:0-99}(a)), the maximum out-of-plane corrugation is $\delta z_{\textrm{KC-QMC}} = 0.075~\textrm{\AA}$.
The maximum corrugation of KC-Ouyang is $\delta z_{\textrm{KC-Ouyang}} = 0.071~\textrm{\AA}$, or 5\% smaller than $\delta z_{\textrm{KC-QMC}}$.
This similarity is expected from the agreement in SFE (Fig.~\ref{fig:gsfe}).
For KC-DFT-D2, the chiral relaxations are also observed between layers, where the top layer has the opposite chirality to the bottom layer, in agreement with KC-QMC and KC-Ouyang. 
However, the difference in maximum out-of-plane corrugation from KC-QMC is more pronounced as $\delta z_{\textrm{KC-DFT-D2}}$ is $0.112~\textrm{\AA}$, which is 50\% larger than $\delta z_{\textrm{KC-QMC}}$.
A small difference from KC-QMC is also observed in around AA nodes, in which an inner 6-fold symmetric structure is observed, while in the case of KC-QMC and KC-Ouyang, the AA regions have smooth round-shaped bulges. 
On the other hand, the structure from KC-DFT-D3 has much smaller overall corrugation, with the maximum of only $\delta z_{\textrm{KC-DFT-D3}} = 0.045~\textrm{\AA}$.
As a result, the chiral symmetry between both layers and 6-fold symmetric structure around AA nodes are not observed in KC-DFT-D3.

\subsection{Effects of the more accurate minimum energy structure on the band structure of TBG}

\begin{table}
    \caption{\label{tab:bandwidth}
    Flat bandwidths in meV of the structures from different potentials at various commensurate twist angles.
    }
    \begin{ruledtabular}
    \begin{tabular}{lrrrrrr}
    Potential &   0.84$^{\circ}$ &   0.93$^{\circ}$ &  0.99$^{\circ}$ &   1.05$^{\circ}$ &   1.08$^{\circ}$ &   1.16$^{\circ}$ \\
    \hline
    KC-QMC & 45.680 & 18.627 & 6.294 & 16.098 & 26.350 & 49.134 \\
    KC-Ouyang & 44.998 & 18.075 & 5.848 & 16.901 & 27.108 & 49.933 \\
    KC-DFT-D2 & 44.218 & 16.938 & 5.880 & 18.974 & 29.404 & 52.534 \\
    KC-DFT-D3 & 42.936 & 17.744 & 4.740 & 16.352 & 26.175 & 48.243 \\
\end{tabular}
\end{ruledtabular}
\end{table}

\begin{figure}
\includegraphics{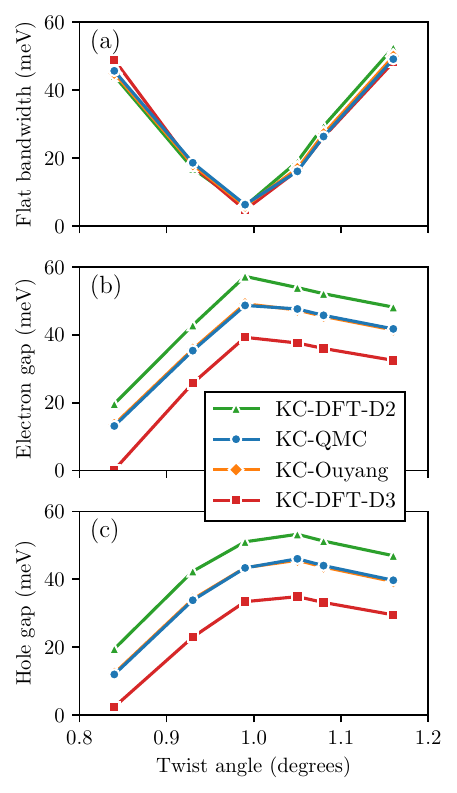}
\caption{\label{fig:gaps} 
(a) Flat bandwidth, (b) electron band gap, (c) hole band gap as a function of twist angle. 
The four sets of band structure are computed using the same LETB model~\cite{pathak2022accurate} but with geometries relaxed under different interlayer KC potentials: KC-QMC, KC-Ouyang, KC-DFT-D2, and KC-DFT-D3.}
\end{figure}

To investigate the effects of the more accurate minimum energy structure on the band structure, the LETB model is employed~\cite{pathak2022accurate}.
In Fig.~\ref{fig:0-99}(e, f, g, h), the LETB band structures are plotted for four twisted bilayer structures at $0.99^{\circ}$, relaxed using KC parameters from four different sources, namely KC-QMC, KC-Ouyang, KC-DFT-D2, and KC-DFT-D3.
The band structures from KC-QMC, KC-Ouyang, and KC-DFT-D3 show similar features, in which the four bands exhibit two degenerate energy states at the gamma point, as reported in previous studies~\cite{carr2018pressure,goodwin2020hartree,pathak2022accurate}.
The band structure from a KC-DFT-D2 geometry shows a slightly different feature from the other three structures, where in this case, all the four bands seem to form a single degenerate energy state at the gamma point.

The flat bandwidths, defined as the difference between the maximum energy and the minimum energy of the flat bands, for different potentials and twist angles are reported in Table~\ref{tab:bandwidth}.
At the twist angle of $0.99^{\circ}$, the bandwidth for KC-QMC is 7\% larger than KC-Ouyang and KC-DFT-D2, and 33\% larger than KC-DFT-D3.
In Fig.~\ref{fig:gaps}(a), the flat bandwidth is plotted as a function of the twist angle. 
Reducing the twist angle from $1.16^{\circ}$ to $0.99^{\circ}$, all four potentials show a similar downward trend, where the inflection points occurs, which defines the first magic twist angle.
The identification of the magic angle at $0.99^{\circ}$ agrees with the previous result~\cite{pathak2022accurate}.

The electron and hole gaps are defined by the separation of the flat bands from the remote bands, and have been noted~\cite{lucignano2019crucial,choi2019electronic} to be particularly sensitive to the corrugation. 
For our geometries, these gaps are presented in Fig.~\ref{fig:gaps}(b, c). 
The treatment of the vdW interaction can change the estimated hole and electron gaps by a significant amount. 
For example, near the magic angle of $0.99^{\circ}$, the gaps can vary by almost a factor of two. 
Corrugation tends to increase the electron and hole gaps, so the underestimation of corrugation in DFT-D3 results in gaps that are too small, while the reverse in DFT-D2 results in gaps that are too large.

\section{Conclusion}

As has been noted previously in the literature, the careful treatment of the van der Waals interaction appears to be very important to obtain accurate corrugation in twisted bilayer graphene~\cite{zhou2015van,yoo2019atomic}, and small changes in the corrugation can affect the electronic structure significantly~\cite{rakib2022corrugation,uchida2014atomic}.
In this work, we provided a state-of-the-art benchmark of the van der Waals interaction in bilayer graphene and found that while a commonly used atomic potential~\cite{ouyang2018nanoserpents} is fairly accurate, DFT-based methods may not improve the description, and in fact may lead to worse results. 
Models of the electronic structure which depend on van Hove singularities~\cite{wu2021chern} and other features of the flat bands should take this into consideration.

For the accuracy of the interlayer interaction in graphene, we found that the commonly used Kolmogorov--Crespi~\cite{kolmogorov2005registry} model as parameterized in Ref.~\cite{ouyang2018nanoserpents} was surprisingly accurate compared to QMC results, as well as the RPA estimation.
On the other hand, commonly used corrections to DFT, the DFT-D2 and DFT-D3 functionals, resulted in large errors in the interlayer potential and stacking fault energy, which leads to an over (DFT-D2) or under (DFT-D3) estimation of the degree of corrugation. 

We found that the electronic band structure varied significantly depending on the treatment of the corrugation. 
While structures relaxed from different potentials result in the same prediction of the magic twist angle at $0.99^{\circ}$, the error in the electronic structure due to using DFT geometries near the magic angle (here found to be roughly $0.99^{\circ}$) results in rearrangements of bands up to around 50\% ($\approx 3~\mathrm{meV}$) in the two lower flat bands for the DFT-D2 structure.
While the separation of the flat bands from the other bands as quantified by electron and hole gaps shows the same inflection point at $0.99^{\circ}$, at the magic twist angle, the relaxed structures from DFT-D2 and DFT-D3 result in a similar error of roughly $10~\mathrm{meV}$, which is an error of about 50\%. 

This study has focused on freestanding bilayer graphene and has employed fixed-node QMC as a reference method due to its established reliability. 
We believe this technique is the most accurate available that can treat the required system size.
Comparison between QMC and less accurate approximation techniques serves as a benchmark for testing and improving numerical schemes, and it would be interesting to see whether other high-accuracy first-principles techniques obtain similar results. 
We only considered freestanding bilayer graphene in this study, while most experiments are performed on a substrate and encapsulated with materials such as boron nitride. 
It would be a fruitful future direction to use this model to explore the effects of encapsulation on the structure of bilayer graphene.

\section{Data Availability}

Alongside this paper, we provide the interlayer energy of bilayer graphene from QMC and a potential fit to the QMC data suitable for use in \lstinline{LAMMPS}, which is available online at \url{https://github.com/qmc-hamm/qmc_graphene_stacking_fault}.
This data is suitable for use to perform atomic scale simulations of bilayer graphene, and has the level of accuracy comparable to the underlying quantum Monte Carlo calculations in the binding region. 

\section{Acknowledgements}
This work was supported by the U.S. Department of Energy, Office of Science, Office of Basic Energy Sciences, Computational Materials Sciences Program, under Award No. DE-SC0020177.
This work made use of the Illinois Campus Cluster, a computing resource that is operated by the Illinois Campus Cluster Program (ICCP) in conjunction with the National Center for Supercomputing Applications (NCSA) and is supported by funds from the University of Illinois at Urbana-Champaign.
This research used resources of the Oak Ridge Leadership Computing Facility at the Oak Ridge National Laboratory, which is supported by the Office of Science of the U.S. Department of Energy under Contract No. DE-AC05-00OR22725.

\section{Appendix}

\subsection{Errors in diffusion quantum Monte Carlo data}

\begin{figure}
    \includegraphics{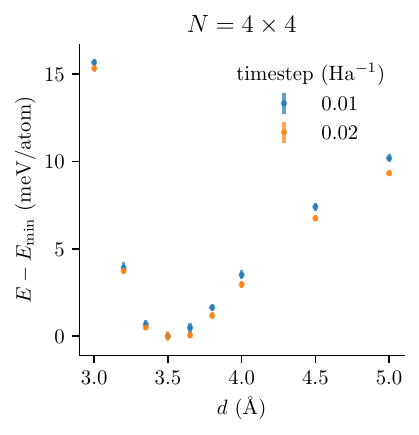}
    \caption{\label{fig:timesteps}
    QMC energy  vs. interlayer distance for AB-stacked bilayer graphene with the supercell size of $N = 4 \times 4$, where $N$ is the number of primitive cells in a simulation cell.
    Energies are referenced to the minimum energy point at 3.5 \AA.
    The energies for the time step of $0.02~\mathrm{Ha^{-1}}$ are roughly 0.01 meV away from the energies for the time step of $0.01~\mathrm{Ha^{-1}}$.
    }
    \end{figure}

Figure~\ref{fig:timesteps} shows that the results for the time step of $0.02~\mathrm{Ha^{-1}}$ have roughly 0.01 to 0.02~meV errors from the energies for the time step of $0.01~\mathrm{Ha^{-1}}$.
Since the discrepancy is small at small interlayer distances, we consider the time step of $0.02~\mathrm{Ha^{-1}}$ a good approximation for describing the potential energy near equilibrium distances. 

\begin{figure}
    \includegraphics{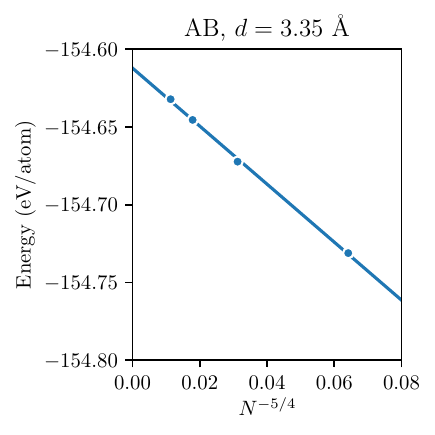}
    \caption{\label{fig:extrap}
    QMC energy vs. $N^{-5/4}$ for AB-stacked bilayer graphene at $3.35~\text{\AA}$ interlayer distance, where $N$ is the number of primitive cells in a simulation cell.
    The statistical error bars from QMC range from 0.2 to 0.8 meV/atom, which are much smaller than the $y$-axis scale and are therefore not displayed in the figure.
    The line is fitted through the energy data points using Eq.~(\ref{eq:extrap}) for the purpose of extrapolating the energy to the thermodynamic limit.
    }
    \end{figure}

The diffusion quantum Monte Carlo (QMC) calculations are performed using a constant $4 \times 4$ twist grid on each of the supercells $N = 3 \times 3$, $4 \times 4$, $5 \times 5$, and $6 \times 6$, where $N$ is the number of unit cells in a simulation cell.
The grid of twists is offset from the $\Gamma$ point by 0.076923 in reciprocal lattice units in order to avoid the Dirac point, which would otherwise result in slower convergence.
The largest cell size results in an equivalent $24 \times 24$ $\mathbf{k}$-mesh, which we confirmed is converged in DFT.
The energy is extrapolated to the infinite system size by fitting the following linear equation~\cite{drummond2008finite}
\begin{align}
E(N) = E(\infty) + cN^{-5/4}, \label{eq:extrap}
\end{align}
where $E(\infty)$ is a fitting parameter, which represents the energy extrapolated to the thermodynamic limit, and $N$ is the number of primitive cells in a simulation cell.
The error bars of the extrapolated energies are calculated using the bootstrapping technique.
The linear model fits our energy data well, as shown in Fig.~\ref{fig:extrap} for AB-stacked bilayer graphene with $3.35~\text{\AA}$ interlayer distance. 
The time step is chosen to be $0.02~\mathrm{Ha}^{-1}$. 

\subsection{Fitting procedure}
We sampled 44 energy data points using QMC and fit them to the KC potential~\cite{kolmogorov2005registry}.
The KC model is a registry-dependent potential designed to improve on the classical potential by taking into account the anisotropy of the $\pi$ overlap between layers.
The full model is given by
\begin{align}
E^{\textrm{KC}} &= \frac{1}{2} \sum_i \sum_{j \neq i} \mathrm{Tap}(r_{ij}) V_{ij}. \label{eq:kc} \\
V_{ij} &= e^{-\lambda(r_{ij} - z_0)} [C + f(\rho_{ij}) + f(\rho_{ji})] - A\left(\frac{r_{ij}}{z_0}\right)^{-6}. \nonumber \\
\rho_{ij}^2 &= r_{ij}^2 - (\mathbf{r}_{ij} \cdot \mathbf{n}_i)^2. \nonumber \\
\rho_{ji}^2 &= r_{ij}^2 - (\mathbf{r}_{ij} \cdot \mathbf{n}_j)^2. \nonumber \\
f(\rho) &= e^{-(\rho/\delta)^2} \sum_{n=0}^2 C_{2n} \left(\frac{\rho}{\delta}\right)^{2n}, \nonumber
\end{align}
where $\mathbf{r}_{ij}$ is the distance vector pointing from atom $i$ to atom $j$ from different layers,
$\mathbf{n}_{k}$ is the surface normal at atom $k$, and $\rho_{ij}$ is the transverse distance from atom~$i$ to atom~$j$.
The taper function given by
\begin{align}
\mathrm{Tap}(x_{ij}) &= 20 x_{ij}^7 - 70  x_{ij}^6 + 84  x_{ij}^5 - 35 x_{ij}^4 + 1, \\
x_{ij} &= \frac{r_{ij}}{R_{\rm{cut}}},
\end{align}
provides a continuous long-range cutoff, and $R_{\mathrm{cut}}$ is fixed to $16~\mathrm{\text{\AA}}$ throughout all the calculations.

The weighted fit is determined by optimizing the function    
\begin{align}
\chi^2 &= \sum_l^{N_{\textrm{d}}} w_l \left(E_l - E_l^{\textrm{KC}} \right)^2, \nonumber
\end{align}
where $E_l$ is the computed energy for the $l$-th configuration of bilayer graphene, $E_l^{\textrm{KC}}$ is the calculated energy from Eq.~(5), and $N_{\textrm{d}}$ is the number of data points, which is 44 for the QMC method and 268 for DFT-D2 and DFT-D3.
The fitting weight $w_l$ depends on the energy and a tunable parameter $k_{\textrm{B}} T$ according to the formula
\begin{align}
w_l &= \exp \left({-\frac{E_l - E_{\mathrm{min}}}{k_\mathrm{B}T}} \right). \label{eq:weights}
\end{align}
Here, $E_{\mathrm{min}}$ is the smallest data point, which in our case is the equilibrium point of the AB stacking potential energy surface, 
$k_{\mathrm{B}}$ is the Boltzmann constant, 
and $T$ is the temperature, which serves as a parameter that controls the relative weight of the low and high energy data points.

The weights are introduced so that we can tune how much the fitting model favors the data points near equilibrium by adjusting the value of $k_{\mathrm{B}} T$.
A smaller value of $k_{\mathrm{B}} T$ corresponds to a model that highly favors the points near equilibrium, while giving up the goodness of fit at very small or large interlayer distances.
Thus, the model is selected according to the goodness of fit in the region of interest, which can be quantified by the root mean square (RMS) error as described in the next section.
The values of fitting parameters for different values of $k_{\mathrm{B}} T$ are reported in Table~\ref{tab:kc_params} along with the KC parameters from Ref.~\cite{ouyang2018nanoserpents} labeled as KC-Ouyang.
In Fig.~\ref{fig:kT}(a, b, c, d), the computed QMC energies for four stacking types are displayed as black data points along with the fitted curves for three different sets of weights, i.e. $k_{\textrm{B}} T = 2, 4$, and $\infty~\mathrm{meV}$ (unweighted).

We find that while the KC potential fits well to the energy data from the DFT-D scheme because they both have the functional form of $r^{-6}$, the QMC potential energy is not exactly proportional to $r^{-6}$, which is why the KC potential struggles to describe the entire potential energy surface as shown in Fig.~\ref{fig:kT}. 
Modifications to the model might be needed to describe the potential energy surface at any interlayer separation range.

\begin{table*}
    \caption{\label{tab:kc_params} 
    List of KC parameter values for the carbon-carbon interaction. 
    The energy within the parentheses after KC-QMC indicate the value of $k_{\textrm{B}} T$, which is used to tune the fitting weights described by Eq.~(\ref{eq:weights}).
    The model KC-QMC (4 meV) are used to describe interlayer interactions to obtain relaxed structures shown in Fig.~\ref{fig:0-99}(a, b, c, d).
    The value of $R_{\text{cut}}$ is fixed to $16~\mathrm{\text{\AA}}$.
    }
    \begin{ruledtabular}
    \begin{tabular}{lllllllll}
    Potential & $z_0~(\mathrm{\text{\AA}})$ & $C_0~(\mathrm{meV})$ & $C_2~(\mathrm{meV})$ & $C_4~(\mathrm{meV})$ & $C~(\mathrm{meV})$ & $\delta~(\mathrm{\text{\AA}})$ & $\lambda~(\mathrm{\text{\AA}^{-1}})$ & $A~(\mathrm{meV})$ \\
    \hline
     KC-QMC (2 meV) & 3.44323 & 14.33525 & 12.56000 & 0.41916 & 7.40186 & 0.65357 & 3.25395 & 13.91535 \\
     KC-QMC (3 meV) & 3.40756 & 16.29563 & 13.09544 & 0.01848 & 6.65729 & 0.68934 & 3.28558 & 13.74788 \\
     KC-QMC (4 meV) & 3.37942 & 18.18467 & 13.39421 & 0.00356 & 6.07494 & 0.71935 & 3.29308 & 13.90678 \\
     KC-QMC (5 meV) & 3.37042 & 19.26299 & 12.89983 & 0.00524 & 5.07182 & 0.75587 & 3.29479 & 13.84456 \\
     KC-QMC (6 meV) & 3.35062 & 21.22680 & 13.17347 & 0.20199 & 4.18306 & 0.78726 & 3.29340 & 14.17078 \\
     KC-QMC (7 meV) & 3.35104 & 22.97568 & 12.37219 & 0.31184 & 1.46736 & 0.86239 & 3.28298 & 14.03609 \\
     KC-QMC (8 meV) & 3.33406 & 25.26089 & 14.07200 & 2.20093 & 0.01279 & 0.84472 & 3.27746 & 14.38600 \\
     KC-QMC (9 meV) & 3.34121 & 24.58252 & 13.29192 & 1.28582 & 0.00060 & 0.86790 & 3.27689 & 14.15974 \\
    KC-QMC (10 meV) & 3.30334 & 27.74635 & 15.07551 & 1.53471 & 0.00001 & 0.86547 & 3.27683 & 15.12378 \\
    KC-QMC (11 meV) & 3.31271 & 26.85873 & 15.25734 & 3.13552 & 0.00087 & 0.82689 & 3.27726 & 14.83406 \\
    KC-QMC (12 meV) & 3.31412 & 26.68834 & 15.26599 & 3.70204 & 0.00005 & 0.81525 & 3.27729 & 14.77050 \\
    KC-QMC (13 meV) & 3.33134 & 25.18924 & 14.44755 & 4.21503 & 0.00008 & 0.80145 & 3.27754 & 14.29565 \\
    KC-QMC (14 meV) & 3.33377 & 24.96005 & 14.28375 & 4.75430 & 0.00000 & 0.79129 & 3.27772 & 14.21445 \\
    KC-QMC (15 meV) & 3.36987 & 22.14523 & 12.65076 & 4.42759 & 0.00341 & 0.78733 & 3.27777 & 13.31070 \\
    KC-QMC (16 meV) & 3.37850 & 21.51294 & 12.19151 & 4.82753 & 0.00005 & 0.77777 & 3.27803 & 13.09407 \\
    KC-QMC (17 meV) & 3.40957 & 19.41387 & 10.93804 & 4.61657 & 0.00002 & 0.77283 & 3.27814 & 12.38366 \\
    KC-QMC (18 meV) & 3.37184 & 21.95300 & 12.29456 & 5.48004 & 0.00011 & 0.76864 & 3.27827 & 13.22834 \\
    KC-QMC (19 meV) & 3.38030 & 21.33727 & 11.89459 & 5.50172 & 0.00005 & 0.76582 & 3.27834 & 13.02214 \\
    KC-QMC (20 meV) & 3.38045 & 21.31321 & 11.82651 & 5.65686 & 0.00002 & 0.76330 & 3.27840 & 13.01085 \\
    KC-QMC (unweighted) & 3.37006 & 21.78334 & 10.46939 & 8.86496 & 0.00001 & 0.72395 & 3.28315 & 13.09016 \\
    KC-Ouyang~\cite{ouyang2018nanoserpents} & 3.41608 & 20.02158 & 10.90551 & 4.27564 & 0.01001 & 0.84471 & 2.93606 & 14.31326 \\

   \end{tabular}
\end{ruledtabular}
\end{table*}

\subsection{Choosing the KC parameter set}

\begin{figure*}
\includegraphics[width=5in]{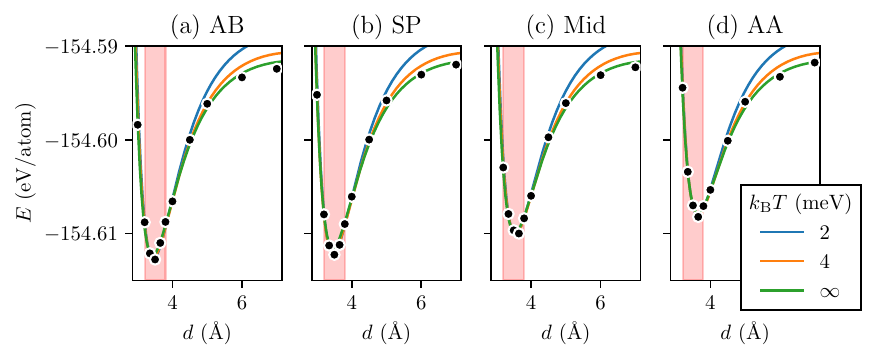}%
\includegraphics[width=2in]{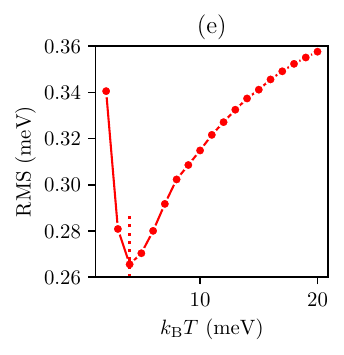}
\caption{
Potential energy surface of bilayer graphene at registry (a) AB, (b) SP, (c) Mid, and (d) AA.
Data points (black) are calculated using QMC.
Lines are fits using the KC model from Eq.~(\ref{eq:kc}) with the fitting weights controlled by the parameter $k_\textrm{B}T$ as given by Eq.~(\ref{eq:weights}).
Red highlighted regions cover the range of interlayer distances near the equilibrium spacing, chosen to be $d \in [3.2, 3.8]~\text{\AA}$.
(e) RMS as a function of $k_{\mathrm{B}} T$ for the data points within the range of interlayer distances $d \in [3.2, 3.8]~\text{\AA}$.
For the region near the equilibrium spacing, $k_{\mathrm{B}} T = 4~\mathrm{meV}$ minimizes RMS (shown as the dotted vertical line).
The KC model associated with this value of $k_{\mathrm{B}} T$ is therefore chosen to represent the QMC data (labeled as KC-QMC) and is used in the subsequent relaxation calculations of the twisted bilayer graphene.
}
\label{fig:kT}
\end{figure*}

\begin{table}
\caption{\label{tab:be}
Binding energy in meV/atom at different registries.
The chosen KC-QMC model corresponds to KC fitting with no weights.
The error bars from KC-QMC are obtained from the bootstrap method.
}
\begin{ruledtabular}
\begin{tabular}{lllll}
Method & AB & SP & Mid & AA \\
\hline
KC-QMC & 21.96(9) & 21.17(8) & 19.52(9) & 17.41(9) \\
KC-Ouyang & 24.82 & 24.11 & 22.28 & 19.92 \\
KC-DFT-D2 & 23.99 & 22.93 & 20.65 & 17.82 \\
KC-DFT-D3 & 19.52 & 19.02 & 17.95 & 16.36 \\
RPA~\cite{olsen2013beyond} & 25 & & & \\
QMC~\cite{mostaani2015quantum} & 17.7(9) & & & 11.5(9) \\
\end{tabular}
\end{ruledtabular}
\end{table}

\begin{table}
\caption{\label{tab:dmin}
The minimum interlayer spacing $d_{\mathrm{min}}$ in $\text{\AA}$.
The chosen KC-QMC model corresponds to $k_{\mathrm{B}} T = 4~\mathrm{meV}$.
The error bars from KC-QMC are obtained from the bootstrap method.
}
\begin{ruledtabular}
\begin{tabular}{lllll}
Method & AB & SP & Mid & AA \\
\hline
KC-QMC &  3.442(4) & 3.460(4) & 3.548(5) & 3.650(6) \\
KC-Ouyang & 3.395 & 3.420 & 3.499 & 3.611 \\
KC-DFT-D2 & 3.260 & 3.290 & 3.389 & 3.531 \\
KC-DFT-D3 & 3.542 & 3.566 & 3.626 & 3.720 \\
RPA~\cite{olsen2013beyond} & 3.39 & & & \\
Exp.~\cite{razado2018structural} & 3.48(10) & & & \\
\end{tabular}
\end{ruledtabular}
\end{table}

In order to decide which KC parameter set to use, we investigate the RMS errors between the QMC data and the fitting within the region of interest.
The RMS errors for the data within $d = 3.2$ to $3.8~\text{\AA}$ as illustrated by the highlighted regions in Fig.~\ref{fig:kT}(a, b, c, d) are plotted as a function of $k_{\mathrm{B}} T$ in Fig.~\ref{fig:kT}(e).
This range of interlayer distances is chosen based upon the fact that the relaxed interlayer distances of all the methods we explore fall within this range as shown in Fig.~\ref{fig:gsfe}(b).
Figure~\ref{fig:kT}(e) suggests that the most preferable model is corresponding to $k_{\mathrm{B}}T = 4~\mathrm{meV}$ due to having the smallest RMS. 
Therefore, all the subsequent calculations of relaxed structures in this work are performed using this set of KC parameters to describe the interlayer interactions. 
In the case of KC-QMC, the chosen parameters are the third row of Table~\ref{tab:kc_params}.
The minimum interlayer spacing $d_{\textrm{min}}$ for different stacking registries and potentials is reported in Table~\ref{tab:dmin}.

On the other hand, the binding energy (BE) is defined as the energy required to separate the two graphene sheets in equilibrium to the infinite interlayer distance.
Therefore, the BE is a property at a large interlayer distance and is evaluated using KC parameters that are fitted with no weights because data points at large interlayer distances now deserve the same weights as data points near equilibrium.
The BEs for different stacking registries and potentials are reported in Table~\ref{tab:be}.
Our BEs from QMC for AB-stacked and AA-stacked bilayer graphene are 21.96(9) meV and 17.41(9) meV, while the BEs for these two stacking configurations are reported to be 17.7(9) and 11.5(9) meV in the previous QMC study~\cite{mostaani2015quantum}, which are slightly smaller than our BE results.
The small discrepancy between the QMC BEs may arise from various factors, including differences in QMC data points, fitting functions, and equilibrium spacing $d_{\textrm{min}}$ being used to estimate the BEs between the two studies (Table~\ref{tab:compare_mostaani}).
We also observed that precise optimization of the Jastrow factor is crucial for obtaining consistent results, with errors similar to the observed discrepancies when the Jastrow factor is not optimized extremely carefully.

\begin{table}
    \caption{\label{tab:compare_mostaani}
    Comparison of key factors that could lead to discrepancy between the QMC results of this study and Ref.~\cite{mostaani2015quantum}.
    }
    \begin{ruledtabular}
    \begin{tabular}{lrr}
        Factor & Present work & Ref.~\cite{mostaani2015quantum} \\
    \hline
Data points & 44 & 5 \\
Fitting potential & KC~\cite{kolmogorov2005registry} & ``16-12-8-4''~\cite{mostaani2015quantum} \\
$d_{\textrm{min}}$ for AA & $3.650(6)~\text{\AA}$ & $3.495~\text{\AA}$~\cite{brihuega2012unraveling} \\
$d_{\textrm{min}}$ for AB & $3.442(4)~\text{\AA}$ & $3.384~\text{\AA}$~\cite{brihuega2012unraveling} \\
Method to obtain $d_{\textrm{min}}$ & QMC & vdW-DF~\cite{rydberg2003van,dion2005erratum,roman2009efficient} \\
    \end{tabular}
\end{ruledtabular}
\end{table}

\begin{table}
    \caption{\label{tab:gsfe_constants}
    Fitting constants in meV for the stacking-fault energy  as a function of registry as given by Eq.~(\ref{eq:gsfe}). 
    The error bars from KC-QMC are obtained from the bootstrap method.
    }
    \begin{ruledtabular}
    \begin{tabular}{lllll}
         Method &        $c_0$ &        $c_1$ &        $c_2$ &        $c_3$ \\
    \hline
KC-DFT-D2 & 2.46701 & -0.71409 & -0.13147 & 0.02322 \\
KC-Ouyang & 1.85743 & -0.56041 & -0.07521 & 0.01648 \\
   KC-QMC & 1.815(38) & -0.546(9) & -0.098(19) & 0.039(8) \\
      RPA & 1.74278 & -0.49974 & -0.09179 & 0.01101 \\
KC-DFT-D3 & 1.15472 & -0.34776 & -0.03980 & 0.00266 \\
    \end{tabular}
\end{ruledtabular}
\end{table}

\begin{table}
    \caption{ Fitting constants in {\AA} for the relaxed interlayer spacing  as a function of registry as given by Eq.~(\ref{eq:gsfe}). 
    The error bars from KC-QMC are obtained from the bootstrap method.
    }
    \begin{ruledtabular}
    \begin{tabular}{lllll}
    Method &        $c_0$ &        $c_1$ &        $c_2$ &        $c_3$ \\
    \hline
KC-DFT-D2 & 3.35457 & -0.03093 & -0.00147 & 0.00073 \\
KC-Ouyang & 3.47106 & -0.02458 & -0.00149 & 0.00056 \\
   KC-QMC & 3.5171(21) & -0.0245(5) & -0.0021(9) & 0.0015(4) \\
      RPA & 3.47805 & -0.02632 & -0.00363 & 0.00039 \\
KC-DFT-D3 & 3.60477 & -0.01990 & -0.00122 & 0.00007 \\
    \end{tabular}
    \end{ruledtabular}
\end{table}

\bibliography{paper}{}

\end{document}